\documentclass[useAMS]{mn2e}
\usepackage{times}
\def\eq#1 {eq.~(\ref{eq:#1})}

\def\etal{{\it et al.\ }}

\def\ie{{\it i.e.},}

\def\ltsima{$\; \buildrel < \over \sim \;$}
\def\lsim{\lower.5ex\hbox{\ltsima}}
\def\gtsima{$\; \buildrel > \over \sim \;$}
\def\gsim{\lower.5ex\hbox{\gtsima}}
\def\ga{\mathrel{\hbox{\rlap{\hbox{\lower4pt\hbox{$\sim$}}}\hbox{$>$}}}}
\def\la{\mathrel{\hbox{\rlap{\hbox{\lower4pt\hbox{$\sim$}}}\hbox{$<$}}}}

\def\log{{\rm log}}

\def\la{\dangler}

\def\pmb#1{\setbox0=\hbox{#1}%
 \kern-.025em\copy0\kern-\wd0
 \kern.05em\copy0\kern-\wd0
 \kern-.025em\raise.0433em\box0}

\def\3hmpc{\, ( h^{-1} {\rm Mpc})^3}

\renewcommand{\H}{{\mbox{${\mathrm{H}}$~{\small I}~}}}
\newcommand{\Hp}{{\mbox{${\mathrm{H}}$~{\small II}~}}}
\newcommand{\He}{{\mbox{${\mathrm{He}}$~{\small I}~}}}
\newcommand{\Hep}{{\mbox{${\mathrm{He}}$~{\small II}~}}}
\newcommand{\Hepp}{{\mbox{${\mathrm{He}}$~{\small III}~}}}

\def\r0p { r{_0^\prime}}

\title[Heating of the Intergalactic Medium by Primordial
Miniquasars]{Heating of the Intergalactic Medium by Primordial
Miniquasars} \author[Zaroubi \etal]{Saleem Zaroubi$^1$, Rajat
M. Thomas$^1$, Naoshi Sugiyama$^2$ \& Joseph Silk$^3$ \\$^1$Kapteyn
Astronomical Institute, University of Groningen, Landleven 12, 9747 AG
Groningen, The Netherlands \\$^2$
Department of Physics and Astrophysics, Nagoya University,
Chikusa-ku Nagoya 464-8602, 
Japan\\ $^3$ Astrophysics Department, University of Oxford, Keble
Road, Oxford OX1 3RH }

\begin{document}
\maketitle
\begin{abstract}
A simple analytical model is used to calculate the X-ray heating of
the IGM for a range of black hole masses. This process is efficient
enough to decouple the spin temperature of the intergalactic medium
from the cosmic microwave background (CMB) temperature and produce a
differential brightness temperature of the order of $\sim
5-20~\mathrm{mK}$ out to distances as large as a few co-moving Mpc,
depending on the redshift, black hole mass and lifetime. We explore
the influence of two types of black holes, those with and without
ionising UV radiation.  The results of the simple analytical model are
compared to those of a full spherically symmetric radiative transfer
code.  Two simple scenarios are proposed for the formation and
evolution of black hole mass density in the Universe.  The first
considers an intermediate mass black hole that form as an end-product
of Population~III stars, whereas the second considers super-massive
black holes that form directly through the collapse of massive halos
with low spin parameter. These scenarios are shown not to violate any
of the observational constraints, yet produce enough X-ray photons to
decouple the spin-temperature from that of the CMB.  This is an
important issue for future high redshift 21~cm observations.

\end{abstract}


\section{Introduction}
\label{introduction}

One of the most startling findings made in the last few years is the
discovery of super-massive black holes at redshifts $\gsim 5.7$ with
black hole masses of the the order of $10^9 M_\odot$ (Fan \etal 2003
and 2006). The origin and seeds of these black holes remain
uncertain.  

Currently, there are two main scenarios for creating such massive
black holes. One is as the end-product of the first metal free stars
(Population~III stars) that have formed through molecular hydrogen
cooling (Abel, Bryan, Norman 2000, 2002, Bromm, Coppi \& Larson 2002,
Yoshida \etal 2003). Given the low cooling rate provided by molecular
hydrogen, the collapsing initial cloud is expected not to be able to
fragment into small masses and thus produce very massive stars (for
reviews, see Bromm \& Larson 2004, Ciardi and Ferrara 2005).  These
stars are expected to burn their fuel very quickly and to produce
black holes with masses in the range $30 - 1000 M_\odot$ (O'Shea and
Norman 2006), with the exception of the mass range of $140 - 260
M_\odot$ where the pair-instability supernovae leave no black hole
remnants (Bond, Arnett \& Carr 1984; Heger \& Woosley 2002; Rakavy,
Shaviv \& Zinamon 1967).  Such objects grew their masses very
efficiently by accretion up to $10^9 M_\odot$ by $z\approx 6$
(Volonteri \& Rees 2005, Rhook \& Haehnelt 2006).

The second avenue for producing even more massive black holes is
through the collapse of very low angular momentum gas in rare
dark-matter halos with virial temperatures above $10^4 \mathrm{K}$
(see Shapiro 2004 for a recent review). Under such conditions, atomic
cooling becomes efficient and black holes with masses $\gg 10^3
M_\odot$ can be formed (Bromm \& Loeb 2003). Fragmentation of the
initial gas into smaller mass objects due to efficient cooling can be
prevented by trapping the Lyman-$\alpha$ photons within the collapsing
gas (Spaans \& Silk 2006).

Notwithstanding the origin of these massive black holes, their impact
on the intergalactic medium is expected to be dramatic in at least two
ways. Firstly, these objects produce very intense ionising radiation
with power-law behaviour that creates a different ionization pattern
around them from that associated with thermal (i.e., stellar)
sources. The ionization aspect of the miniquasar radiation has been
explored by several authors (Madau, Meiksen \& Rees 1997, Ricotti \&
Ostriker 2004a, 2004b, Madau \etal 2004, Zaroubi \& Silk
2005). Recently, however, it has been argued (Dijkstra, Haiman \& Loeb
2004, Salvaterra, Haardt \& Ferrara 2005) that miniquasars can not
reionize the Universe as they will produce far more soft X-ray
background radiation than currently observed (Moretti et al. 2003;
So{\l}tan 2003) and at the same time satisfy the WMAP 3$^{rd}$ year
polarisation results (Page \etal 2006 ; Spergel \etal 2006) and the
reionization constraints from the IGM temperature at redshift $\approx
3$ (Theuns \etal 2002a, 2002b, Schaye \etal 2000). It should be noted,
however, that the Dijkstra \etal 2004 \& Salvaterra \etal\ 2005
calculations have been carried out assuming specific black-hole mass
density evolution histories and spectral energy distributions of
UV/X-ray radiation emanating from the miniquasars.

Secondly, due to their x-ray radiation, even the intermediate mass
black holes (IMBH) are very efficient in heating up their
surroundings.  Nusser (2005) has pointed out that this heating
facilitates observation of the redshifted 21 cm radiation in either
emission or absorption by the neutral hydrogen in the high-redshift
IGM. The observation of this radiation is controlled by the 21 cm spin
temperature, $T_{spin}$, defined through the equation $n_1/n_0=3
\exp(-T_\ast/T_{spin})$. Here $n_1$ and $n_0$ are the number densities
of electrons in the triplet and singlet states of the hyperfine
levels, and $T_\ast=0.0681~\mathrm{K}$ is the temperature
corresponding to the 21 cm wavelength. For the 21 cm radiation to be
observed relative to the CMB background, it has to attain a different
temperature and therefore must be decoupled from the CMB (Wouthuysen
1952; Field 1958, 1959; Hogan \& Rees 1979). The decoupling is
achieved through either Lyman$-\alpha$ radiation or collisional
excitations and heating. For the objects we are concerned with in this
paper, \ie\ miniquasars, the collisional excitation and heating are
much more important.  In general, throughout this paper, we will
ignore the influence of Lyman-$\alpha$ photons emitted by the quasar
on $T_{spin}$. However, one should point out that collisional
excitations due to x-ray photons results in a ``secondary''
Lyman$-alpha$ pumping which will dominate the spin temperature and CMB
temperature decoupling in some regions around the miniquasar; this
effect has been recently point out by Chuzhoy, Alvarzez \& Shapiro
(2006).  For recent papers that discuss X-ray heating, see Chen \&
Miralda-Escude (2006) and Pritchard \& Furlanetto (2006).

Collisional decoupling of $T_{spin}$ from $T_{CMB}$ is caused by very
energetic electrons released by the effect of the x-ray miniquasar
radiation on the IGM. Shull \& van Steenberg (1985) have estimated
that more than a tenth of the energy of the incident photons is
absorbed by the surrounding medium as heating (this fraction increases
rapidly with the ionized fraction). The increase in the temperature is
observable at radio frequencies in terms of the differential
brightness temperature, $\delta T_b$, which measures the 21 cm
intensity relative to the CMB. A similar fraction of the absorbed
energy also goes into collisional excitation, where this fraction
decreases rapidly with the ionised fraction. These two processes,
heating and excitation decouple the spin temperature from the CMB
temperature and render the IGM observable through its 21 cm emission.

Recently, Kuhlen \& Madau (2005) and Kuhlen, Madau \& Montgomery
(2006) have performed a detailed numerical study of the influence of
150$M_\odot$ IMBH on its surroundings and calculated the gas, spin and
brightness temperatures. They have shown that heating by 150$M_\odot$
IMBH at $z=17.5$ can enhance the 21 cm emission from the warm neutral
IGM. The filaments enhance the signal even further and may make the IGM
visible in  future radio experiments (e.g., the LOFAR-Epoch of
Reionization key science project\footnote{For more details on the
LOFAR radio telescope see http://www.lofar.org}).

In this paper, we adopt a complementary theoretical approach to the
numerical one adopted by Kuhlen \& Madau (2005). This allows us to
explore the influence of power-law radiation fields from a range of
black hole masses that are presumed to reside in the centres of
primordial miniquasars. Furthermore, the effect of X-ray induced
collisional excitations on the 21 cm spin temperature is included
(Chuzhoy, Alvarez \& Shapiro 2006) -- this effect is not taken into
account in the Kuhlen \& Madau (2005) work. We test two main classes
of x-ray emitting miniquasars, those with UV ionizing radiation and
those without.  We show that in both cases these miniquasars might
play an important role in heating the IGM without necessarily ionizing
it completely.

In addition, two simple scenarios for the formation of (mini)quasars
as a function of redshift are presented. This is done using the
extended Press-Schechter algorithm to predict the number density of
forming black holes either with H$_2$ cooling or with atomic
cooling. We also discuss the implications of these scenarios for the
mass density of quasars at redshift~6, the soft X-ray background
(SXRB)in the energy range $0.5-2~\mathrm{keV}$ (Dijkstra, Haiman \&
Loeb 2004), the number of ionizing photons per baryon and, finally,
the optical depth for Thomson scattering of CMB photons.

The paper is organised as follows: Section~\ref{Theory} describes the
theoretical methods used here and derives the ionization and kinetic
temperature profiles around miniquasars without UV ionizing
radiation. Section~\ref{Tspin} calculates the spin and brightness
temperature around the same quasars. In section~\ref{IMBH} we show the
ionization and heating profiles around quasars with UV ionizing
radiation.  Section~\ref{PS} present the two formation scenarios and
their implications. The paper concludes with a summary section
(\S~\ref{summary}).

\section{Heating and kinetic temperature}
\label{Theory}

The exact shape of the UV and X-ray photon spectral energy
distribution around high redshift mini-quasars is uncertain. However,
in general it is believed to have two continuum components. The first
is through to emanate from the putative accretion disk around a
black hole; this component, at least in low mass black holes, is well
described by ``multicolour disk blackbody''(Mitsuda \etal 1984).  The
hottest blackbody temperature, $T_{max}$, in a Keplerian disk damping
material onto a black hole at the Eddington rate is $k T_{max}\approx
k1 \mathrm{keV} (M/M_\odot)^{-1/4}$ (Shakura \& Sunyaev 1973), where
the hole mass, $M$, is measured in solar mass units. The
characteristic multicolour disk spectrum follows a power law with $L_E
1 \propto E^{1/3}$ at $E< kT_{max}$. The second component spectrum is
a simple power law with spectral energy distribution proportional to
$E^{-\alpha}$ with $\alpha approx 1$. The precise origin of this power
law is uncertain and very likely to be due to nonthermal processes.

To simplify the calculation, we follow Kuhlen \etal 2005 and consider
miniquasars with power-law flux spectra and power-law index of
$-1$. We also assume, at this stage, that the ionizing UV photons
produced by the miniquasars are absorbed by the immediate black hole
environment. Therefore a lower cutoff of the photon energies is
assumed, namely,
\begin{equation}
F(E)= {\cal A} E^{-1}~\mathrm{s^{-1}} \ \ \ \ \ \ \{ 200~\mathrm{eV}
\le E \le 100~\mathrm{keV} \}.
\label{spectrum}
\end{equation}
where ${\cal A}$ is normalized such that the miniquasar luminosity is
a tenth of the Eddington luminosity. Miniquasars with UV ionizing
photons are considered in  a later stage in this paper.

This spectrum translates to a number of photons per
unit time per unit area at distance $r$ from the source,
\begin{equation} 
{\cal N}(E;r) = e^{-\tau(E;r)} \frac{{\cal A}}{\left(4 \pi
r^2\right)}
E^{-1}\mathrm{cm^{-2} s^{-1}},
\label{flux}
\end{equation}
with
\begin{equation} 
\tau(E;r)=\int_0^r n_H x_{HI} \sigma(E) dr \, .
\label{tau}
\end{equation}

Here $x_{H I}$ is the hydrogen neutral fraction, $n_H \approx
1.9\times 10^{-7}~\mathrm{cm^{-3}} (1+z)^3$ (Spergel \etal 2006) is the
mean number density of hydrogen at a given redshift, and $\sigma_H(E)
= \sigma_0 \left({E_0/E}\right)^3$ is the bound-free absorption
cross-section for hydrogen with $\sigma_0=6\times
10^{-18}~\mathrm{cm^2}$ and $E_0=13.6~eV$.  The second equation is obtained assuming a
homogeneous density for the IGM.

The cross-section quoted earlier does not take into account the
presence of helium. In order to include the effect of helium, we
follow Silk \etal (1972) who modified the cross section to become
\begin{equation}
\sigma(E) = \sigma_H(E) +{n_{He} \over n_H} \sigma_{He} = \sigma_1
\left({E_0 \over E}\right)^3.
\end{equation}
A proper treatment of the effect of helium is accounted for by
defining $\sigma_1$ to be a step function at the two helium ionisation
energies corresponding to He I and He II. This however includes
lengthy calculations and complicates the treatment, and we therefore
choose $\sigma_1$ to be a smooth function of $E,$ an approximation
that will overestimate $\sigma(E)$ for low energy photons. For the
kinds of spectra and energies we consider here, this is a reasonable
assumption.

\subsection{Ionization}

To obtain the optical depth at a given distance, $r$, from the
miniquasar, we calculate the neutral fraction around the miniquasar
for a given spectrum and energy range by solving the
ionization-recombination equilibrium equation (Zaroubi \& Silk 2005):
\begin{equation}
\alpha_{H I}^{(2)} n_H^2 (1-x_{H I})^2= \Gamma(r)~n_H x_{H I}
\left(1+{\sigma_{He}\over\sigma_{H} }{n_{He}\over n_{H}}\right).
\label{balance}
\end{equation}
Here $\Gamma(r)$ is the ionisation rate per hydrogen atom at distance
$r$ from the source. Since we are interested in the detailed structure
of the ionisation front, $\Gamma$ is calculated separately for each
value of $r$ using the expression,
\begin{equation}
\Gamma (r) = \int_{E_0}^{\infty}\sigma(E) {\cal N}(E;r)\times \left( 1 + \frac{E}{E_0}~\phi(E,x_{e})\right)\frac{dE}{E}.
\end{equation}
The function $\phi(E,x_e)$ is the fraction of the initial photon
energy that is used for secondary ionizations by the ejected electrons
and $x_e$ is the fraction of ionized hydrogen (Shull \& van Steenberg
1985; Dijkstra, Haiman \& Loeb 2004). The
$\frac{E}{E_0}~\phi(E,x_{e})$ term is introduced to account for the
number of ionization introduced by secondary ionization.  Furthermore,
in eq.~\ref{balance} $\alpha_{H I}^{(2)}$ is the recombination
cross-section to the second excited atomic level and has the values of
$2.6 \times 10^{-13}T_4^{-0.85} \mathrm{cm^3 s^{-1}}$, with $T_4$
being the gas temperature in units of $10^{4~} \rm K$. For this
calculation we assume that $T=10^{4~}\mathrm{K}$. This is of course
not very accurate, although it gives a lower limit on the
recombination cross-section, $\alpha_{H I}^{(2)}$ (in neutral regions
atomic cooling prevents the gas from having a higher
temperature). Since the region we are going to explore is mostly
neutral, an accurate estimation of the recombination cross-section is
not necessary.
%
%
\begin{figure}
\setlength{\unitlength}{1cm} \centering
\begin{picture}(8,4.8)
\put(-1.,-.5){\includegraphics{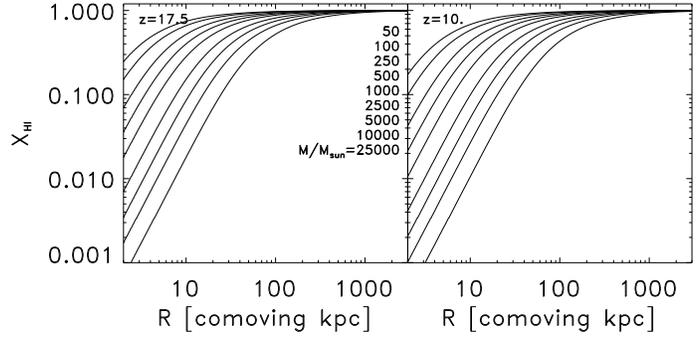}}
\end{picture}
\caption{The neutral hydrogen fraction as a function of distance for a
range of black hole masses for z=17.5 (left) and z=10 (right) for
miniquasars without ionising UV radiation, namely, with radiation that
spans the energy range of $200 \mathrm{eV}< E <10^5\mathrm{eV}$}.
\label{fig:xneut1}
\end{figure}

Figure~\ref{fig:xneut1} shows the solution of equation~\ref{balance}
for miniquasars with masses ranging from $50 M_\odot$ up to $2.5\times
10^4 M_\odot$. We assume that the miniquasars emit at a tenth of the
Eddington luminosity and that their emitted radiation is confined to
$200\le E\le 10^5 \mathrm{eV}$. The lack of ionising UV photons
results in a very small ionized region around the miniquasar centres
(x-ray photons are not very efficient in ionization) with an extended
transitional region between the ionised and the neutral IGM (Zaroubi
\& Silk 2005). We also assume that the density of the IGM around the
miniquasars is the mean density in the Universe (this could be easily
replaced by any spherical density profile). Due to the increase of the
mass density at higher redshifts, the ionising photons are absorbed
closer to the quasar. The neutral fraction profile obtained for each
profile is used in the following sections to calculate the kinetic,
spin and brightness temperatures of the IGM surrounding the
miniquasars.

\subsection{Heating}
The heating rate per unit volume per unit time that is produced by the
photons absorbed by the IGM for a given photon energy at distance $r$
from the source is ${\cal H}(r)$. ${\cal H}$ is calculated separately
for each $r$ using the expression,
\begin{equation}
{\cal H}(r) = f n_{H} x_{HI}(r)\int_{E_0}^{\infty}\sigma(E) {\cal N}(E;r)
dE
\label{eq:heat}
\end{equation}
where $f$ is the fraction of the absorbed photon energy that goes into
heating through collisional excitations of the surrounding material
(Shull \& van Steenberg 1985). The function $f$ is fitted in the Shull
and van Steenberg (1985) paper with the following simple fitting
formula: $f = C \left[ 1- \left( 1- x^a\right)^b \right]$, where
$C=0.9771$, $a=0.2663,$ $b=1.3163$ and $x=1-x_{HI}$ is the ionized
fraction. This fitting function is valid in the limit of high photon
energies, an appropriate assumption for the case at hand. We only
modify the fitting formula by imposing a lower limit of $11\%$ for the
fraction of energy that goes into heating as the proposed fitting
formula does not work well at ionized hydrogen fractions smaller than
$10^{-4}$. This equation is similar to that obtained by Madau, Meiksin
\& Rees (1997).
%
%
\begin{figure}
\setlength{\unitlength}{1cm} \centering
\begin{picture}(8,9.4)
\put(-1.,-0.5){\includegraphics{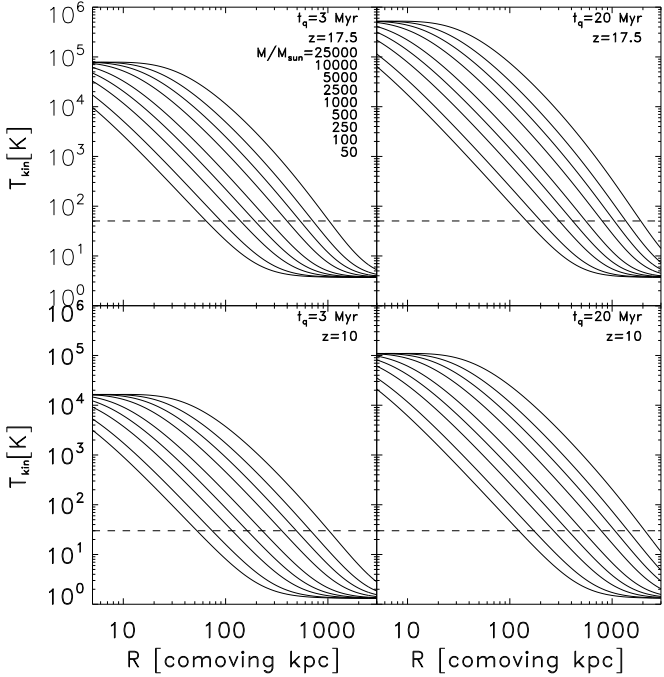}}
\end{picture}
\caption{The kinetic temperature of the gas for a range of black hole
masses. The redshift and quasar lifetime ($t_q$) are specified on each
panel. The dashed line indicates the CMB temperature at the
corresponding redshift.}
\label{fig:tkin}
\end{figure}

In order to determine the temperature of the IGM due to this heating,
we adopt the following equation,
\begin{equation}
 \frac{3}{2}\frac{n_H k_b T_{kin}(r)}{\mu} = {\cal H}(r) \times t_q.
\end{equation}
Here, $T_{kin}$ is the gas temperature due to heating by collisional
processes, $k_b$ is the Boltzmann constant, $\mu$ is the mean
molecular weight and $t_q$ is the miniquasar lifetime. This equation
assumes that the heating rate due to the absorption of x-ray photons
during the miniquasar lifetime is constant. Given the miniquasar
lifetime relative to the age of the Universe at the redshifts we are
interested in, cooling due to the expansion of the Universe can be
safely neglected. Notice that in the highly ionized regions, although
we ignore it, Compton cooling off CMB photons for long living
miniquasars and high redshifts should be included.

Figure~\ref{fig:tkin} shows the kinetic temperature as a function of
radius for the same black hole masses considered in
figure~\ref{fig:xneut1}. The heating of the IGM is clearly very
extended and ranges from about a quarter of a comoving Mpc for a
black hole with $50~M_\odot$ up to more than $3$ comoving Mpc for
black holes with masses $\gsim 10^4~M_\odot$. Since the mass density
in the Universe increases towards higher $z$ as $(1+z)^3$, the neutral
fraction around the miniquasar is larger, hence, the heating is more
effective at higher redshifts. The figure also shows that, as
expected, the heating is larger for a quasar with a longer lifetime.
Note, that at redshift 10 other effects (e.g, Lyman-$\alpha$ pumping,
metal cooling lines) might play a more important role than heating by
miniquasars. However, the purpose of presenting the $z=10$ figures is
to show the redshift trend of change due to miniquasars.

%
%
\begin{figure}
\setlength{\unitlength}{1cm} \centering
\begin{picture}(8,9.4)
\put(-1.,-0.5){\includegraphics{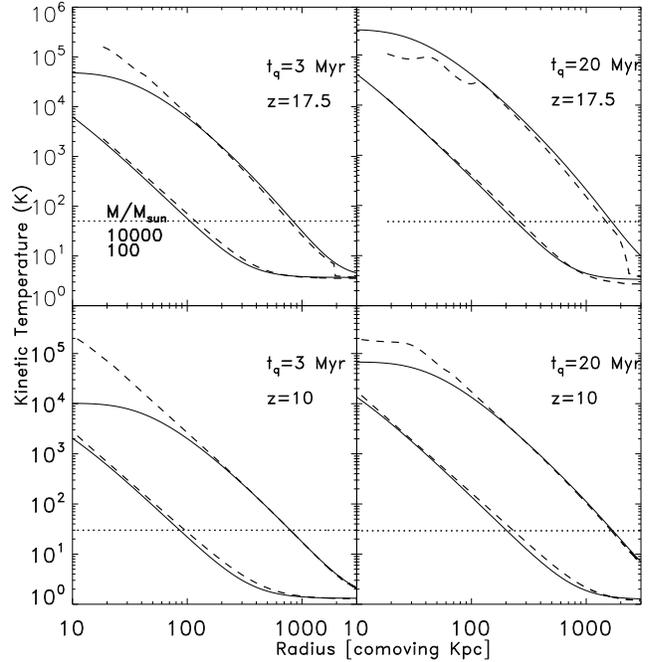}}
\end{picture}
\caption{This figure shows a comparison between the model adopted in
this study and the results from a spherically symmetric radiative
transfer code (Thomas \& Zaroubi 2006) applied to two of the IMBH
masses, 100 \& 10000 $M_\odot$ with the same radiation power spectrum.
The analytical calculation is represented by the solid line and that
obtained from the radiative transfer code is represented by the dashed
line. The dotted line indicates the CMB temperature at the
corresponding redshift.}
\label{fig:compare}
\end{figure}
\subsection{Comparison with a spherically symmetric full radiative transfer code}

In order to test our analytical approach we compare our results with
those obtained by running a non-equilibrium spherically symmetric
radiative transfer code that is applied to the same problem. Details
of the code are described by Thomas \& Zaroubi (2006) but here we give
a brief description. The radiative transfer code evolves
non-equilibrium equations for \H, \Hp, \He, \Hep, \Hepp, $\mathrm{e}$
and the electron temperature $T_{e}$. The equations take into account
collisional and photo-ionization, recombination, collisional
excitation cooling, recombination cooling, free-free cooling, Hubble
cooling, Compton heating and Compton cooling. The comparison between
the analytical and the numerical results is performed for 8 cases. The
8 cases constitute all combinations of two black hole masses (100 and
10000), two redshifts (10 and 17.5) and two miniquasar lifetimes (3
and 20 Mega-years). The comparison is shown in
figure~\ref{fig:compare} where the kinetic temperature of the gas
obtained from the simple analytical calculation is represented by the
solid line and that obtained from the radiative transfer code is
represented by the dashed line. Except at the centre where the neutral
fraction adopted profile differs in the two cases, the agreement
between the two approaches is very good. The main reason for the
departure in the centre is that the equilibrium solution assumes that
the neutral fraction profile shown in fig~\ref{fig:xneut1} is attained
within the quasar lifetime; this assumption is simply incorrect for
high energy photons where the bound-free time scales exceeds that (for
a recipe to mitigate this effect see Thomas \& Zaroubi 2006). To
summarise, given the many processes included in the radiative transfer
code, this agreement is satisfactory.

Another comparison one can make is with the gas temperatures obtained
by Kuhlen \& Madau (2005) shown in the upper right panel of figure~7
in their paper. Visual inspection of the results of our approach when
applied to a 150 $M_\odot$ IMBH with the same spectrum shows good
agreement. Both of these comparisons give us confidence in the validity of the
simplistic theoretical approach adopted in this paper.

\section{21-cm spin and brightness temperatures}
\label{Tspin}

\subsection{The Spin Temperature}

In his seminal paper, Field (1958; see also Kuhlen, Madau \&
Montgomery 2006) used the quasi-static approximation to calculate the
spin temperature, $T_{spin},$ as a weighted average of the CMB
temperature, the gas kinetic temperature and the ``light'' temperature
related to the existence of ambient Lyman-$\alpha$ photons (Wouthuysen
1952; Field 1958). The spin temperature
is given by:
\begin{equation}
T_{spin} = \frac{T_\ast + T_{CMB} +y_{kin} T_{kin} + y_\alpha
T_{kin}}{1+y_{kin}+y_\alpha},
\label{eq:tspin}
\end{equation}
where $T_{CMB}$ is the CMB temperature and $y_{kin}$ and $y_\alpha$
are the kinetic and Lyman-$\alpha$ coupling terms, respectively.

The Kinetic coupling term is due to the increase in the kinetic
temperature due to X-ray heating.
\begin{equation}
y_{kin}=\frac{T_\ast}{A_{10} T_{kin}}\left(C_H+C_e+C_p \right).
\end{equation}
Here $A_{10}=2.85\times 10^{-15}~\mathrm{s^{-1}}$ (Wild 1952) is the
Einstein spontaneous emission rate coefficient.  $C_H$, $C_e$ and
$C_p$ are the de-excitation rates due to neutral hydrogen, electrons
and protons, respectively. These rates have been calculated by several
authors (Field 1958; Smith 1966; Allison \& Dalgarno 1969; Zygelman
2005). In this paper we use the fitting formulae used in Kuhlen, Madau
\& Montgomery (2006) which we repeat here for completeness, the rate
due to neutral hydrogen $C_H=3.1\times 10^{-11} n_H T_{kin}^{0.357}
\exp(-32/T_{kin}) \rm [s^{-1}]$; the rate due to electrons is $C_e=n_e
\gamma_e$ where $\log(\gamma_e/1~\mathrm{cm^3~s^{-1}}) = -9.607 +
0.5~\log~T_{kin} \times \exp\left( -(\log~T_{kin})^{4.5}/1800\right)$;
and the rate due to protons is $C_p=3.2~n_p~\kappa$, where
$\kappa=C_H/n_H$ is the effective single-atom rate coefficient. And
$n_H$, $n_e$ and $n_p$ are the hydrogen, electron and proton number
densities in the unit of $\rm cm^{-3}$, respectively and $T_{kin}$ is
measured in K.

The Lyman-$\alpha$ coupling term is also due to collisional
excitation. Previously, studies that have considered x-ray heating
have ignored this effect. Recently however, Chuzhoy \etal\ (2006) have
pointed out that this contribution is very important and even
dominates the spin-temperature value in a certain temperature
range. In order to account for this term one should calculate the
intensity of the Lyman-$\alpha$ photons due to collisional
excitations, $J_0$. This is given by the following equation:
\begin{equation}
{J_0}(r) = \frac{\phi_\alpha\,c}{4 \pi H(z) \nu_\alpha} n_{H}
x_{HI}(r)\int_{E_0}^{\infty}\sigma(E) {\cal N}(E;r) dE.
\label{eq:excite}
\end{equation}
This equation is similar to eq~\ref{eq:heat} except that instead of
the fraction of the absorbed energy that goes to heat, $f$, one should
use the fraction of the absorbed energy that goes into kinetic
excitation of Lyman-$\alpha$. The fraction, $\phi_\alpha$, is also
parameterised by Shull \& van Steenberg (1985) and is given by,
$\phi_\alpha \approx 0.48 \left( 1- x^{0.27}\right)^{1.52}$ (where
$x=1-x_{HI}$). In the equation above, $c$ is the speed of light and
$\nu_\alpha$ is the Lyman-$\alpha$ transition frequency. The Hubble
constant as a function of redshift, $H(z)$, is calculated assuming
$\Omega_m=0.24$ and $\Omega_\Lambda=0.76$.

The $y_\alpha$ coupling term is (Field 1958),
\begin{equation}
y_\alpha= \frac{16 \pi^2 T_\ast e^2 f_{12} J_0}{27 A_{10} T_{kin} m_e
c}.
\end{equation}
Here, $f_{12}=0.416$ is the oscillator strength of the Lyman$_\alpha$
transition, $A_{10}$ is the Einstein spontaneous emission coefficient
of the 21 cm transition and $e$ \& $m_e$ are the electron charge and
mass, respectively.

%

%
%
\begin{figure}
\setlength{\unitlength}{1cm} \centering
\begin{picture}(8,9.4)
\put(-1.,-0.5){\includegraphics{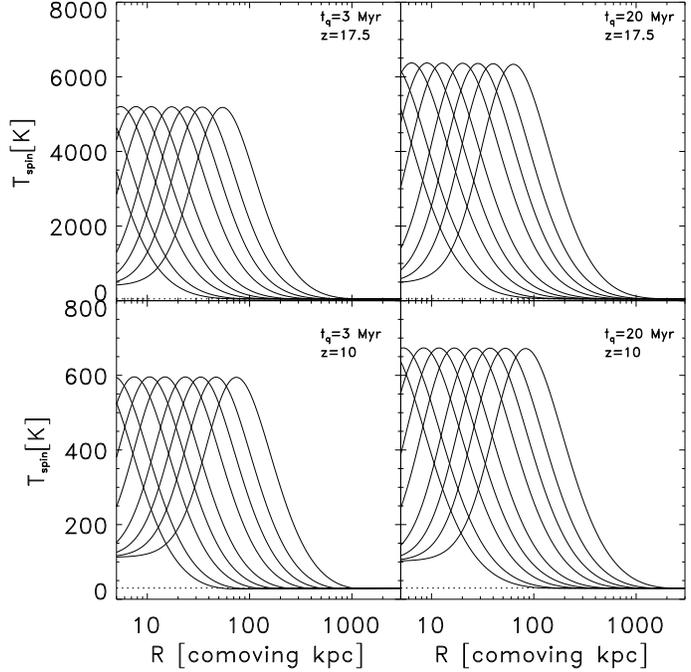}}
\end{picture}
\caption{The spin temperature of the gas for a range of black hole
masses. The redshift and quasar lifetime ($t_q$) are specified on each
panel.  The dotted line indicates the CMB temperature at the
corresponding redshift. Note the different y-axis range between the
$z=17.5$ and $z=10$ panels.}
\label{fig:stemp}
\end{figure}

Figure~\ref{fig:stemp} shows the spin temperature of the gas for a
range of black hole masses. The redshift and quasar lifetime ($t_q$)
are specified on each panel. The figure clearly shows that as the
distance from the miniquasar increases, the temperature drops to the
$T_{CMB}$ level. The distance at which the temperature reaches the
$T_{CMB}$ asymptotic value depends on the black hole mass. For the
more massive black holes, this distance can exceed a couple of
comoving Mpc.

The figure also shows that the maximum spin temperature is almost
independent of the quasar mass -- a detailed inspection of the figure
shows a slight change in the maximum of $T_{spin}$ as a function of
mass. This effect is due to the fact that the dominant coupling
parameter in equation~\ref{eq:tspin} around the maximum $T_{spin}$ is
$y_\alpha$ (by at least an order of magnitude) and is of the order of
$0.01$. Under such conditions equation~\ref{eq:tspin} reduces to
$T_{spin} \approx y_\alpha T_{kin}$, namely, $T_{spin}\propto
J_0$. $J_0$ in regions where $x_{HI}\ll 1$ is independent of quasar
mass as implied by the left hand side of
equation~\ref{balance}. Physically, this means that when the medium is
already ionized no additional heating of the IGM due to bound-free
absorption is possible, no matter how much radiation comes out of the
quasar.

\subsection{The Brightness Temperature}

In radio astronomy, where the Rayleigh-Jeans law is usually
applicable, the radiation intensity, $I(\nu)$ is expressed in terms of
the brightness temperature, so that
\begin{equation} 
I(\nu) = \frac{2 \nu^2}{c^2} k_b T_b,
\end{equation}
where $\nu$ is the radiation frequency, $c$ is the speed of light and
$k$ is Boltzmann's constant (Rybicki \& Lightman 1979). This in turn
can only be detected differentially as a deviation from $T_{CMB}$, the
cosmic microwave background temperature. The predicted differential
brightness temperature deviation from the cosmic microwave background
radiation, at the mean density, is given by (Field 1958, 1959; Ciardi
\& Madau 2003),
\begin{eqnarray}
\delta T_b & = &
\left(20~\mathrm{mK}\right)\left(1+\delta\right)\left(\frac{x_{HI}}{h}\right)\left(1-\frac{T_{CMB}}{T_{spin}}\right)
\nonumber \\ & &\times \left(\frac{\Omega_b h^2}{0.0223}\right)
\left[\left(\frac{1 +
z}{10}\right)\left(\frac{0.24}{\Omega_m}\right)\right]^{1/2},
\end{eqnarray}
where $h$ is the Hubble constant in units of $100
\mathrm{km~s^{-1}~Mpc^{-1}}$, $\delta$ is the mass density contrast,
and $\Omega_m$ and $\Omega_b$ are the mass and baryon densities in
units of the critical density. We also adopt a standard model universe
with a flat geometry, $\Omega_b h^2 =0.022$, $\Omega_m=0.24$ and
$\Omega_\Lambda=0.76$ (Spergel \etal 2006).

\begin{figure}
\setlength{\unitlength}{1cm} \centering
\begin{picture}(8,9.4)
\put(-1.,-0.5){\includegraphics{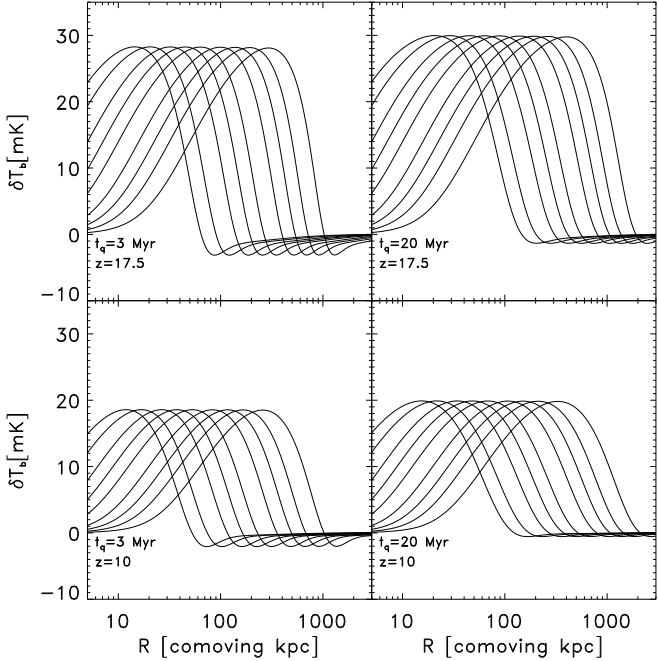}}
\end{picture}
\caption{The brightness temperature for the same cases shown in
figure~\ref{fig:stemp}.}
\label{fig:dTb}
\end{figure}

Figure~\ref{fig:dTb} shows the brightness temperature for the same
IMBH mass explored in figure~\ref{fig:tkin}. The curves show that the
radius at which the differential brightness temperature is detectable
increases with the black hole mass and the miniquasar lifetime
(lefthand vs. righthand panels). The maximum amplitude, however, does
not depend on the black hole mass and depends only weakly on the
miniquasar lifetime. This is because at the centre, $T_{spin} \gg
T_{CMB}.$ Hence $\delta T_b$ is at its maximum value which, at the
mean density of the Universe, only depends on the redshift and
cosmological parameters.

%
%

\section{Miniquasars with ionising UV radiation}
\label{IMBH}

%
%

\begin{figure}
\setlength{\unitlength}{1cm} \centering
\begin{picture}(8,4.8)
\put(-1.,-0.5){\includegraphics{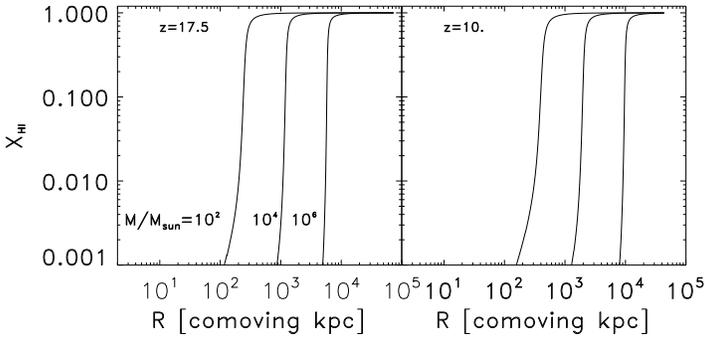}}
\end{picture}
\caption{The neutral hydrogen fraction as a function of distance for 3
black hole masses ($100$, $10^4$ and $10^6~M\odot$) for z=17.5 (left)
and z=10 (right) for miniquasars with UV ionization energy, \ie\
emitted radiation that spans the energy range $10.4~\mathrm{eV} -
\sim10^4~\mathrm{eV}$.}
\label{fig:xneut2}
\end{figure}

We consider the signature of (mini-)quasars with UV radiation that
ionizes the IGM. The different options for quasar spectral energy
distribution has been discussed earlier. Here we follow Madau \etal
2004 and assume that the radiation flux spectrum is the same as in
equation~\ref{spectrum}, except that the energy spans the range of
$10.4~\mathrm{eV}$-$100~\mathrm{keV}$. Of course in this case the
quasar will ionise its immediate surroundings and heat up a more
extended region of the IGM, a realistic spectrum will probably be
between this case and the previous case of truncated power-law (see
\S~ref{PS} for a more complex energy spectrum). Here we test three
black hole masses of $100$, $10^4$ and $10^6~M\odot$ at $z=10$ and
$17.5$ with lifetimes of $3\mathrm{Myr}$. The $10^6~M_\odot$ mass
objects could be considered as progenitors of the SDSS $z\approx 6$
quasars. The \H neutral fraction as a function of distance from the
quasar is shown in figure~\ref{fig:xneut2} for the three black hole
masses at $z=17.5$ (left) and $z=10$ (right).

\begin{figure}
\setlength{\unitlength}{1cm} \centering
\begin{picture}(8,4.8)
\put(-1.5,-0.5){\includegraphics{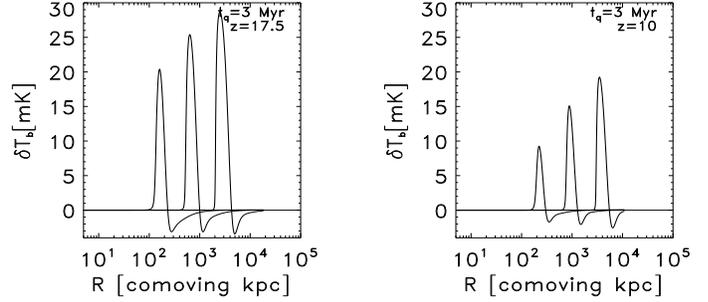}}
\end{picture}
\caption{The differential brightness temperature for 3 miniquasars
with black hole masses $100$, $10^4$ and $10^6~M_\odot$ and ionizing
UV and X-ray photons (\ie\ energy range of $10.4 \mathrm{eV}< E
<10^4\mathrm{eV}$). The quasar lifetime here is 3 Myr. }
\label{fig:dTb_loE}
\end{figure}

If one assumes that the IGM is not heated relative to the CMB, then
the quasar will heat its environment but appears as an emission shell
around the quasar in the 21~cm brightness temperature
maps. Figure~\ref{fig:dTb_loE} shows the differential brightness
temperature around the same three black hole masses shown in
figure~\ref{fig:xneut2}.  The clear difference in the brightness
temperature between this figure and figure~\ref{fig:dTb} is due to the
size of the ionized region around the (mini-)quasar.
%
%

\section{Quasar formation and evolution}
\label{PS}

\subsection{Quasar evolution with redshift}
  
In this section we propose two very simple scenarios for the
production and evolution of quasars at high redshift and explore the
implications for IGM heating, ionization and the observed x-ray
background (XRB) (Moretti et al. 2003, Soltan et al. 2003). We
evaluate the initial mass density of black holes as a function of
redshift, without mass accretion, with the following formation
scenarios: (i)- black holes as end products of stars that have formed
through molecular hydrogen cooling, i.e., stars formed in halos with
virial temperatures smaller than $10^4$~K. (ii)- black holes that have
been produced directly through the collapse of massive low angular
momentum halos. In both cases, we use the Press-Schechter (Press \&
Schechter 1974) formalism with the Sheth \& Tormen (1999) mass
function to infer the number density of halos with a given mass as a
function of redshift.

\begin{figure}
\setlength{\unitlength}{1cm} \centering
\begin{picture}(8,14)
\put(-3.,-1.7){\includegraphics{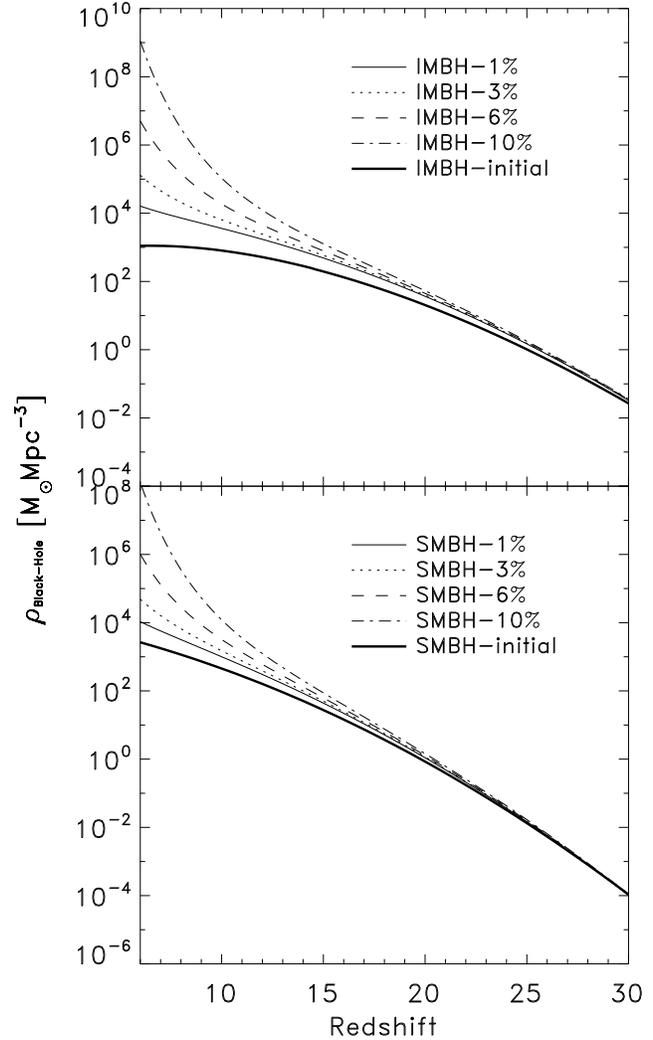}}
\end{picture}
\caption{Initial and evolving comoving black hole mass density as a
function of redshift. The solid thick line shows the mass density of
forming black holes as a function of redshift. The other 4 lines show
the total comoving mass-density for 4 values of $f_{duty}$.  The IMBH
results are shown in the upper panel and those for the SMBH case are shown in
the lower panel.}
\label{fig:PS}
\end{figure}

The mass density of black holes for the first scenario is estimated
simply by calculating the number density of the most massive halos
with molecular hydrogen cooling. These are halos in the range of $0.1
M_{T_4}\le M \le M_{T_4}$, where $M_{T_4}$ is the mass of a halo with
virial temperature $10^{4~}\mathrm{K}$. This is a rough approximation
for halos that have efficient self-shielding for H$_2$ disassociation
and can form pop~III stars through molecular hydrogen cooling (Haiman,
Rees \& Loeb 1997a, 1997b). We henceforth refer to this scenario as
the intermediate mass black hole (IMBH) scenario. To estimate the
comoving mass density of the forming black holes as a function of
redshift, we assume that at the centre of these massive halos, the
star ends its life as a $100~M_\odot \times
\left({M_{halo}/M_{T_4}}\right)$ black hole. The mass density of the
forming black holes as a function of redshift is presented by the
thick solid line shown in the upper panel of figure~\ref{fig:PS}.

For the second scenario, we estimate the number of halos with atomic
hydrogen cooling, namely halos with virial temperature $T_{virial} \ge
10^{4~}\mathrm{K}$. In order to estimate the comoving mass density of
black holes per comoving Mpc$^3$ produced by this scenario, we assume
that only 1\% of the halos in this mass range have a low enough spin
parameter to allow a direct collapse of the halo to form a massive
black hole. The distribution of the spin parameter of halos is quite
flat at the low end of the possible spin parameter range (Steinmetz \&
Bartelmann 1995), and therefore, the choice of 1\% is rather
conservative. In these halos, we take the mass that ends up in black
holes as $10^{-3}\times\frac{\Omega_b}{\Omega_m} M_{halo}$, where the
$10^{-3}$ reflects the Magorian relation between the halo mass and
black hole mass, and $\frac{\Omega_b}{\Omega_m}$ gives the baryon
ratio. The comoving mass density of black holes produced in this type
of scenario is presented by the solid thick line shown in the lower
panel of figure~\ref{fig:PS}. We refer to this model as the
super-massive black hole (SMBH) scenario.

To calculate the accumulated comoving black hole mass-density at any
redshift, we assume that the black hole is accreting at the Eddington
rate with a given radiative efficiency, $\epsilon_{rad}$. The
radiative efficiency is fixed in this paper to be $10\%$. The
cumulative comoving mass density is then given by the following
equation,
\begin{equation}
\tilde\rho(z) = \int\limits_{z}^{35} dz' \rho(z')
e^{f_{duty}\left(\frac{t(z)-t(z')}{t_E}\right)\frac{1-\epsilon_{rad}}{\epsilon_{rad}}}
\; [M_\odot/\rm Mpc^3],
\end{equation}
where $f_{duty}$ is the duty cycle, which ranges from $1\%$ to $10\%$,
$t(z)$ is the age of the universe at redshift $z$ and $t_E \equiv 0.41
\rm Gyr$ is the Eddington time-scale.

The thin lines shown in figure~\ref{fig:PS} show the comoving
black hole mass density as a function of redshift for several
$f_{duty}$ values.  The calculation is done for both IMBH and SMBH
scenarios. The case with $f_{duty}=10\%$ produces a black hole density
relative to the critical density of $\Omega_{black hole}(z=6) \sim
10^{-3}$ and $10^{-4}$ for the IMBH and SMBH scenarios,
respectively. These values are too high to be compatible with the
inferred black hole density at redshift 6. The other extreme case with
$f_{duty}=1\%$ produces $\Omega_{black hole}(z=6) \sim 10^{-8}$ for
both scenarios, which is too low. Therefore, in the following
subsections, we will focus on the results obtained from the cases with
$f_{duty}=3\%$ and $6\%$.

Recently, Begelman, Volonteri \& Rees (2006) have estimated build up
of the black hole mass density at the high redshift Universe that form
via the 'bars within bars' mechanism. This mechanism allows for
the super-massive black holes to form directly in the nuclei of
protogalaxies, without the need for 'seed' black holes left over from
early star formation. In their paper, Begeleman \etal (2006) shown
black hole density as a function of redshift for two duty duty cycles,
$f_{duty}=0.1$ and $0.5$. Unlike in our simple model, their model
gives a rapid increase in the mass density of black holes until
$z\approx 18$ after which the black hole density evolves relatively
slowly (roughly as $\log\rho_{black~hole} = const. + 0.086 z - 0.009
z^2$, which is obtained from a cubic spline fit to their figure 2). To
summarise, according to Begeleman \etal the black hole density attains
relatively high values early on but evolves slowly afterwords,
whereas our model the initial density is low but the mass evolution is
more rapid.

Whichever the actual scenario of the evolution of black holes mass
density in the Universe, it is clear that the mass densities obtained
at high redshift contribute significantly to heating the IGM and
decouple the spin temperature from the CMB temperature (see
figs~\ref{fig:dTb}\&~\ref{fig:dTb_loE}).  In the next section, we explore
which of the scenarios we explore is consistent with the currently
available observational constraints.

\subsection{The Soft X-ray Background (SXRB) Constraint}

Recently Dijkstra \etal (2004) and Salvaterra \etal (2005) have shown
that it is very unlikely that miniquasars have ionized the Universe
without violating the observed SXRB luminosity in the energy range
$0.5-2~\mathrm{keV}$ (Moretti \etal\ 2003). In both cases the authors
have assumed a specific black hole mass history -- instantaneous in
the case of Dijkstra \etal and more gradual in the case of Salvaterra
\etal (2005). Our aim here is to check whether the specific black
hole evolution histories proposed in the current study violate this
observational constraint, regardless of whether they ionise the
Universe or not.

It will be shown that our adopted quasar duty cycle, limited from above
by the Soltan \etal (2003) constraint, yields a diffuse x-ray flux
that is consistent with the SXRB constraint.  We assume a mean
reionization history of the Universe according to which the IGM
underwent a sudden reionization at redshift 6. This assumption is
insensitive to our computed SXRB flux, and is conservative, in that it
provides an upper limit on the ionizing flux from (mini-)quasars.  The
SXRB is calculated for various quasar spectrum templates. The purpose
here is twofold. Firstly, to exclude from our models those cases that
violate the SXRB constraints. Secondly, to explore the influence of
various spectral dependences on the SXRB.

The first template is the one we used for the quasars that have no UV
radiation,
\begin{equation}
F(E) = {\cal A}\;E^{-\alpha} \,\,\, \,\,\, 200 \mathrm{eV} < E < 100
\mathrm{keV},
\end{equation}
where the calculation is made for a range of power-law indices,
$\alpha = 2$~--~$0$.  This represents the case in which all the
ionizing radiation is absorbed in the immediate vicinity of the
quasar. The case we explored previously for the heating and ionisation
fronts was specifically for $\alpha=1$.

The second template, which we have also used before, represents the
case in which all the UV radiation escapes the quasar's immediate
surroundings into the IGM. The template used here is:
\begin{equation}
F(E) = {\cal A}\;E^{-\alpha} \,\,\, \,\,\, 10.4~\mathrm{eV} < E < 100
~\mathrm{keV},
\end{equation}
where $\alpha$ spans the same range as before.

The third case we explore is the one with the template introduced by
Sazonov et al. (2004) and has the form,

\[ \mbox{F(E)} = \left\{ 
\begin{array}{ll}
  \mbox{${\cal A}\; E^{-1.7}$ } & \mbox{if $\;\; 10.4 \mathrm{eV} < E
< 1 \mathrm{keV} $} ;\\ \mbox{${\cal A}\; E^{-\alpha}$} & \mbox{if
$\;\; 1 \mathrm{keV} < E < 100 \mathrm{keV}$}; \\ \mbox{${\cal A}\;
E^{-1.6}$} & \mbox{if $\;\; E > 100 \mathrm{keV} $}.
\end{array} 
\right. \] Notice here that we keep the power-law index of the middle
range, $\alpha$, as the varying parameter. The reason is that quasars
in the redshift range $6-10$ with a Sazonov et al. type spectrum
contribute to the observed SXRB mainly in the energy range
$0.5-2~\mathrm{keV}$.

To proceed, we normalise the above equation with respect to the
product of the Eddington luminosity and the radiation efficiency,
$\epsilon_{rad}$. This should be done at a given distance, $r$, from
the quasar which we choose arbitrarily to be 1 Mpc.

A quasar of mass $M$ shines at $\epsilon_{rad}$ times the Eddington
luminosity, namely
\begin{equation}
L_{Edd}(M) = 1.38\times10^{38} \left(\frac{M}{M_\odot}\right) [\rm erg
\; s^{-1}].
\end{equation}
Therefore ${\cal A}$ is given by:
\begin{equation}
 {\cal A}(M) =\frac{ \epsilon_{rad}\; L_{Edd}(M)}{\int
\limits_{E_{range}} E^{-\alpha}\;dE \times 4\pi r^2 } \;[\rm
erg^{\alpha}\;s^{-1}cm^{-2}] \, ,
\end{equation}
where $E_{range} = 10.4~\mathrm{eV}\, - \, 100~\mathrm{keV} $.

In order to calculate the SXRB, we follow Dijkstra \etal\ (2004).  The
contribution of the soft x-ray background observed in the range $0.5
\mathrm{keV} < E < 2 \mathrm{keV} $, given by:
\begin{eqnarray}
\mathrm{SXRB} & = & \left(\frac{\pi}{180}\right)^2\int\limits_{6}^{35}
dz\,d_A(z)^2 \frac{{\cal A}(\tilde\rho(z))}{(d_L(z)/Mpc)^2}\times
\nonumber\\& & \int\limits_{0.5(1+z)}^{2(1+z)}
E^{-\alpha} \; e^{-\tau(E;z)}dE \; \mathrm{[{erg\,s^{-1}cm^{-2}deg^{-2}}.]}
\end{eqnarray}
In the above equation, $\tau(E;z)$ represents the optical depth,
\begin{eqnarray}
\tau(E;z_Q) & = & \frac{c}{H_o \sqrt{\Omega_m}} \int \limits_6^{z_q}
\frac{dz}{(1+z)^{(5/2)}}\nonumber\\ \times & &
[n_{H~I}(z)\sigma_{H~I}(E')+n_{He~I}(z)\sigma_{He~I}(E')],
\end{eqnarray}
where $E'=E(1+z)/(1+z_Q)$, $z_Q$ is the quasar formation redshift,
$n_{H~I}(z)=n_{H~I}(0)\;(1+z)^3$ and $n_{He~I}(z)=n_{He1}(0)\;(1+z)^3$
are the physical density of hydrogen and helium with $n_{H~I}(0)=
1.9\times 10^{-7}\, \rm cm^{-3}$ and $n_{He~I}(0)= 1.5
\times10^{-8}\, \rm cm^{-3}$.
The luminosity distance, $d_L(z)$, to the black hole is calculated
from the fitting formula given by Pen (1999) and $d_A$ is the angular
diameter distance,
\begin{equation}
d_A(z) = \frac{d_L(z)}{(1+z)^2} \, .
\end{equation}
The division by $d_L^2$ accounts for the dimming of the quasar,
whereas the multiplication by $(\pi/180)^2 \times d_A^2$ calculates
the flux received in a one degree$^2$ field of view.  Moreover, the
normalization factor is now made with respect to the mass density of
black holes, and hence it carries an extra $\rm Mpc^{-3}$ in our
units.

\begin{figure}
\setlength{\unitlength}{1cm} \centering
\begin{picture}(8,9)
\put(-1.,-1.){\includegraphics{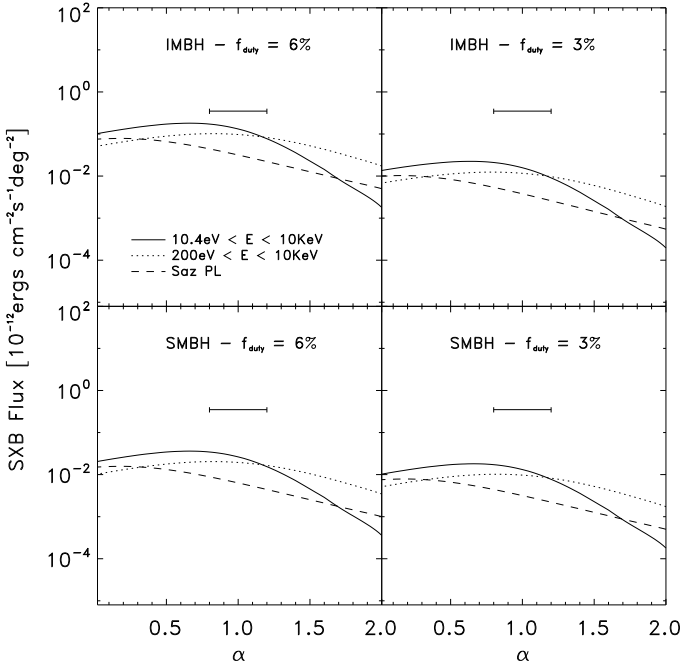}}
\end{picture}
\caption{Soft X-ray background for various spectra.  The four panels
show the SXRB level expected from the IMBH (upper panels) and SMBH
(lower panels) scenarios with $f_{duty}=6\%$ (right panels) and
$3\%$. Each panel shows the SXRB obtained assuming the three
templates: power-law quasars with ionization by UV radiation
(solid-lines) and without UV radiation (dotted-lines) and quasars with
the Sazonov \etal\ 2004 template (dashed-lines).  The short horizontal
line in the middle of each panel marks the observational constraint of
Moretti \etal\ 2003.  }
\label{fig:SXRB63}
\end{figure}

Figure~\ref{fig:SXRB63} shows the expected SXRB as a function of
$\alpha$ for the IMBH and SMBH scenarios in  the $f_{duty}=3\%$ and $6\%$
cases. The short horizontal line at the middle of each of the panels
marks the observational SXRB constraint. This shows that none of these
models violate the observational constraint. The $10\%$ case, which is
not shown here, violates the observed constraints for almost all the
$\alpha$ range.

\subsection{The number of ionizing photons per baryon}
\label{ionize_photon}

We now calculate the number of ionizing photons per baryon emitted in
the IMBH and SMBH models for the $f_{duty}=6\%$ and $3\%$ models.  The purpose of
this calculation is to show that these models will not be able to
ionize the Universe, except in the extreme case in which the escape
fraction of the ionizing UV photons is unity and no recombinations
take place. To estimate the number of ionizing photons, one should
integrate the number of emitted photons per unit energy over the
energy spectrum of the quasars.  The factor $(1-e^{-\tau)}$ accounts
for the absorbed fraction of photons. It also involves an integral
over the active lifetime of the quasars down to redshift 6. These
integrations have the following form:
\begin{eqnarray}
N_{photons} & = & 4\pi \; \int \limits_{6<z<35} dz \; {\cal
A}(\tilde\rho(z))\; \frac{dt}{dz} \; f_{duty} \nonumber \\ &
&\int\limits_{E_{range}} E^{-\alpha} \; (1-e^{-\tau(E;z)})\frac{dE}{E}
\; [\rm Mpc^{-3}],
\end{eqnarray}

where $dt/dz$ is given by,

\begin{equation}
\frac{dt}{dz}= \frac{1}{H_o\;(1+z)\sqrt{(1+z)^2(1+\Omega_m z) -
z(2+z)\Omega_\Lambda}}\; \; [\rm s].
\end{equation}
Again, the mass density parameter $\Omega_m=0.27$ and the vacuum energy
density parameter $\Omega_\Lambda=0.73$.
\begin{figure}
\setlength{\unitlength}{1cm} \centering
\begin{picture}(8,9)
\put(-1.,-1.){\includegraphics{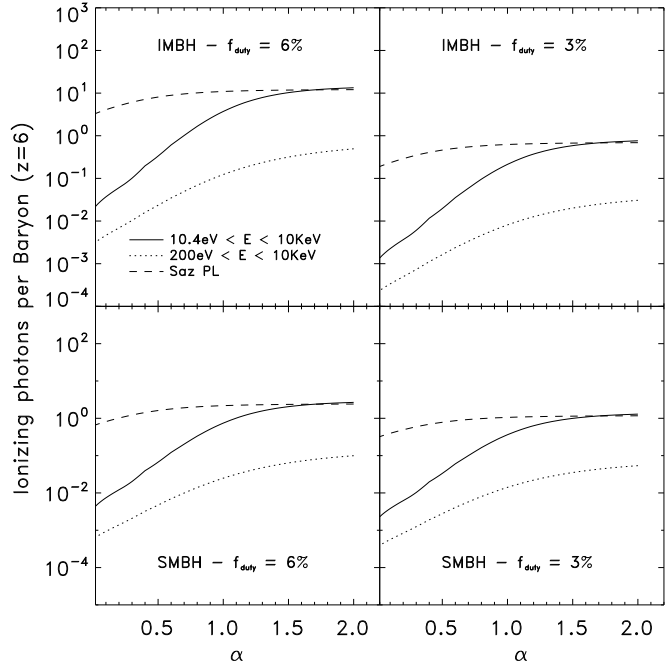}}
\end{picture}
\caption{Number of ionizing photons per baryon for different
spectra. The upper two panels show results for the IMBH scenario with
the left and right hand panels assuming $f_{duty}=6\%$ and $3\%$,
respectively.  The lower two panels show results for the SMBH scenario
with the left and right panels assuming $f_{duty}$ of $6\%$ and
$3\%$. The three models explored are as in the previous figure.}
\label{fig:photon63}
\end{figure}
Figure~\ref{fig:photon63} shows the number of photons per baryon as a
function of the energy spectrum power-law index, $\alpha$. Here we
note a number of features. Firstly, the maximum number of ionizing
photons per baryon is roughly 10. This number is achieved in the IMBH
scenario with $f_{duty}=6\%$ for the spectral templates of both
Sazonov {\it et al.} (dashed line) and the power-law spectrum with
ionizing UV radiation (solid line). Despite obtaining such a high
number of ionizing photons per baryon, one should note that these two
cases assume that all the quasar ionizing photons escape its immediate
surroundings. Not surprisingly, the power-law model without ionizing
photons does not produce too many ionizations (dotted line). Note also
that the number of ionizing photons per baryon produced by the Sazonov
{\it et al.} model does not vary much with $\alpha$. This is simply
because the power-law index we vary in this model is in the X-ray
energy range.

\begin{figure}
\setlength{\unitlength}{1cm} \centering
\begin{picture}(8,9)
\put(-1.,-1.){\includegraphics{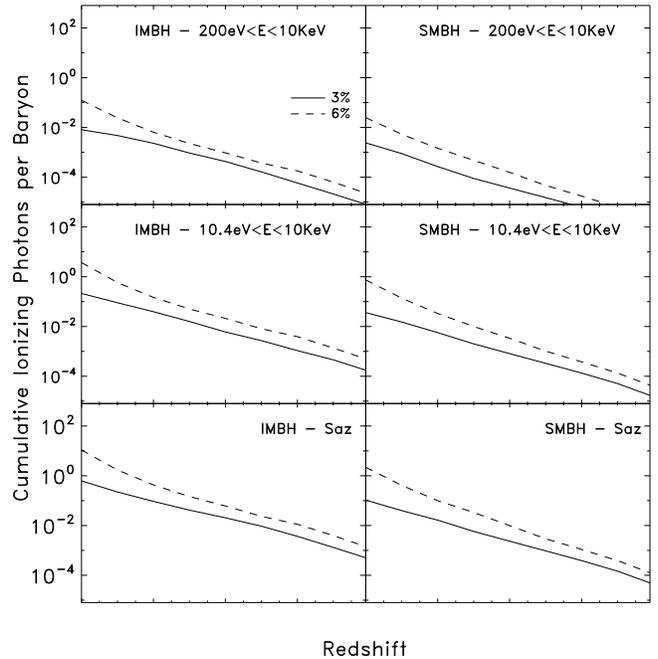}}
\end{picture}
\caption{The number of ionizing photons per baryon as a function of
redshift. The 3 left panels refer to the IMBH scenario with each of
the three showing the number of ionizing photons per baryon for a
different spectral template. The 3 right panels show the same for the
SMBH scenario. These figures assume a power-law index $\alpha$ of 1.
}
\label{fig:cum_photons}
\end{figure}

Figure~\ref{fig:cum_photons} shows the evolution of the number of
ionizing photons per baryon with redshift. The calculation shown here
assumes $\alpha=1$ for all three templates. The 3 left panels show
results for the IMBH scenario, where each of the spectral scenarios is
shown in a different panel. The right hand panels show the same for
the SMBH case. As expected, most of the ionizing photons are produced
towards the low redshift range. The Sazonov {\it et al.} model
produces the largest number of ionizing photons due to its steepness
in the low energy range (power-law index of -1.7). Note, that the
$f_{duty}=6\%$ case produces about 10 photons per baryon at $z=6$
normally thought to be enough to ionize the Universe and the same time
does not violate the SXRB constraint. However, this model does not
reproduce the Thomson $\tau$ constraint (see below).

Assuming that these curves give the actual ionization history, one can
easily calculate the optical depth for Thomson scattering of CMB
photons, $\tau_{CMB}$. This of course is not a self-consistent
calculation since in order to obtain the number of ionising photons as
a function of redshift, one has to assume an ionization history.  This
exercise is still of interest as it gives an upper limit for the
influence of quasars on $\tau_{CMB}$. To calculate $\tau_{CMB},$ we
assume that the electron density, $n_e= n_{H I} \times x_e$ and $x_e$
is given by one-tenth of the number of ionising photons per baryon
that appears in figure~\ref{fig:cum_photons} with an upper limit of
unity. The $\tau_{CMB}$ found here for the IMBH model with
$f_{duty}=6\%$ is about $0.03$ which is only one third of the WMAP3
observed value (Page et al 2006).

\subsection{The predicted constraints from the Begelman \etal (2006) model}

As an aside, we calculate the SXRB, number of ionising photons per
baryons and the Thomson scattering optical depth for the Begelman \etal
(2006) model. Here we choose the case of $f_{duty}=0.1$ and a rigid
disk model, namely, the dotted-line in the lower set of models in their
figure 2. Since their calculation stops at $z=10$ we extrapolate their
black hole mass density curves using a cubic-spline down to $z=6$. For
a Sazonov \etal (2004) type of energy spectrum we obtain SXRB that is
about two orders of magnitude lower than the observed one, which is not
surprising given that the higher the redshift of the miniquasar is, the
less its soft x-ray photons contribute to the observed background. The
number of ionising photons per baryon we obtain is about 10, usually
thought to be the minimum number needed to ionise. However, more
surprisingly we obtain an optical depth for Thomson scattering of,
$\tau\approx 0.075$ which is consistent with the WMAP 3$^{rd}$ year
results (Spergel \etal 2006, Page \etal 2006). This result is
interesting as it shows that for a given black hole evolution history
one can satisfy all the observational constraints and ionise the
Universe solely with black holes at the same time.

\section{summary}
This paper explores the feasibility of heating the IGM with quasars
without violating the current observational constraints. Such heating
is essential in order to be able to observe the 21 cm emission from
neutral hydrogen, prior to and during the epoch of reionization. We
have shown that miniquasars with moderate black hole masses can heat
the surrounding IGM out to radii of a few comoving mega-parsecs.

In this paper, two Press-Schechter based black hole mass density
evolution scenarios have been proposed, IMBH and SMBH. The first model
assumes the black hole population is the end product of pop~III stars
that leave behind black hole masses of the order of 10-100$M_\odot$.
The second model assumes direct formation of black holes as a result
of the collapse of low angular momentum primordial halos.  For these
two scenarios, we have explored three different quasar spectral
templates: a power-law with ionization UV radiation, a power-law
without ionising UV radiation and a Sazonov et al. (2004) type
template.

With the exception of the models that have a $10\%$ duty cycle, we
have shown that the quasars are not able to fully ionise the IGM --
especially if one assumes the template that does not have ionising UV
photons -- while the SXRB constraint is satisfied. We conclude that
based on the mass evolution history shown here, there is enough mass
in the quasars to heat up the IGM by redshift 15. For example, for
quasars with a power-law index of $-1$ and no ionising UV radiation,
quasars with black hole masses of $10^{3-4}M_\odot$ can heat up the
IGM over a $\approx 0.1-1 \mathrm{Mpc}$ comoving radius from the
(mini-)quasar (see figure~\ref{fig:dTb}). The models with $6\%$ duty
cycle reach such mass per comoving Mpc$^3$ at redshift larger than 10
for both scenarios.

Curiously, for the black hole mass density evolution, with $10\%$ duty
cycle and rigid disk model, proposed by Begelman \etal (2006) we find
that this scenario does not violate the SXRB observational constraint
and produce about 10 ionising photons per baryon by $z=6$, normally
thought to be enough to ionise the Universe. We also find that for a
model in which the Universe ionizes suddenly at z=6, that this
scenario predicts a Thomson scattering optical depth of $0.075$,
consistent with the WMAP 3$^{rd}$ year results.

The main result presented in this paper is ``good news'' for the new
generation of low frequency radio telescopes designed to probe the
high redshift IGM through its 21~cm emission, such as LOFAR, MWA and
PAST. It clearly shows that the quasar population could easily
decouple the spin-temperature from that of the CMB.

However, since the spin temperatures achieved are not very high, this
means that the brightness temperature will carry the signature not
only of the ionized fraction and density fluctuations, but also of the
variations in the spin temperature. This complicates the
interpretation of the observed brightness temperature in terms of its
link to the cosmological fields. Nevertheless, the high spin
temperature bubbles are expected to overlap before those of the
ionization, a factor that will mitigate this
complication. Furthermore, one can turn this around and argue that
these fluctuations will teach us more about the ionising sources than
about cosmology. An extended tail in the spin temperature will be a
clear signature of power-law radiation, {\it i.e.}, quasars, while a
short tail will be a clear signature of thermal radiation, {\it i.e.},
stars.

\label{summary}

\section*{acknowledgments}
J. S. \& N.S. acknowledge the Kapteyn Astronomical Institute and
N.S. \& S.Z. acknowledge the Department of Astrophysics at Oxford for
hospitality. The authors thank A. Ferrara and L.V.E. Koopmans for
discussions, A. Nusser for comments on the manuscript and L. Chuzhoy
for drawing our attention to the influence of collisional excitations
on the spin-temperature. The authors thank the anonymous referee for
very helpful comments. N.S. is supported by a Grant-in-Aid for
Scientific Research from the Japanese Ministry of Education
(No. 17540276).

{}
 
\end{document}